\documentclass[preprints,article,accept,moreauthors,pdftex]{mdpi}

\firstpage{1}
\pubvolume{1}
\issuenum{1}
\articlenumber{0}
\pubyear{2021}
\copyrightyear{2020}
\datereceived{}
\dateaccepted{}
\datepublished{}
\hreflink{https://doi.org/} % If needed use \linebreak
\Title{High-precision Time-Frequency Signal Simultaneous Transfer System via a WDM-based Fiber Link}

\TitleCitation{High-precision Time-Frequency Signal Simultaneous Transfer System via a WDM-based Fiber Link}

\Author{Zang Qi $^{1,2,3,5}$,Quan Honglei $^{1,2,3}$,Zhao Kan $^{1}$,Zhang Xiang $^{1,2,3}$,Deng Xue $^{1,2}$,Xue Wenxiang $^{1,2}$,Chen Faxi $^{4}$,Liu Tao $^{1,2,3}$,\\Dong Ruifang$^{1,2,3,*}$ and Zhang Shougang $^{1,2,3}$}

\AuthorNames{Zang Qi, Quan Honglei and Liu Tao}

\AuthorCitation{Qi, Z.; Honglei, Q.; Tao, L.}
\address{%
$^{1}$ \quad National Time Service Center, Chinese Academy of Sciences, 3 shuyuandong Road Xi'an, China., 710600\\
$^{2}$ \quad Key Laboratory of Time and Frequency Primary Standards, Chinese Academy of Sciences, 3 shuyuandong Road Xi'an, China., 710600\\
$^{3}$ \quad University of Chinese Academy of Sciences, 19 yuquan road Beijing, China, 100049\\
$^{4}$ \quad Xidian University, 2 taibainan Road Xi'an, China, 710071\\
$^{5}$ \quad State Key Laboratory of Transient Optics and Photonics, Chinese Academy of Sciences, xinxi Road Xi'an, China, 710119}
\corres{*Correspondence: dongruifang@ntsc.ac.cn}

\abstract {In this paper, we demonstrate a wavelength division multiplexing (WDM) based system for simultaneously delivering ultrastable optical frequency reference, 10 GHz microwave frequency reference, and one pulse per second (1 PPS) time signal via a 50 km fiber network. For each signal, a unique noise cancellation technique is used to maintain the precision of them. After being compensated, the transfer frequency instability in terms of overlapping Allan deviation (OADEV) for the optical frequency achieves 2$\times10^{-17}$/s and scales down to 2$\times10^{-20}$/10000 s,  which for the 10 GHz microwave reference approaches 4$\times10^{-15}$/s and decreases to 1.4$\times10^{-17}$/10000 s, and the time uncertainty of the 1 PPS time signal along the system is 2.08 ps. In this scheme, specific channels of WDM are respectively occupied for different signals to avoid the possible crosstalk interference effect between the transmitted reference signals. To estimate the performance of the above scheme, independent of these signals is also demonstrated in this 50 km link,  the results are similar to that in the case of simultaneous delivery. This work shows that the WDM-based system is a promising method for building a nationwide time and frequency fiber transfer system with a communication optical network.}

\keyword{time and frequency transfer; WDM network; fiber transfer link; transfer ADEV; }

\begin{document}
%%%%%%%%%%%%%%%%%%%%%%%%%%%%%%%%%%%%%%%%%%

\section{Introduction}
With the rapid development of the modern timescales and frequency domain, the frequency instability of the state of the art optical lattice clocks has achieved the $10^{-19}$ level and of $10^{-16}$ for cesium fountain has been demonstrated\cite{ref1,ref2,ref3,ref4,cs,ref5,ref6}. Which are widely used in many advanced scientific and industrial applications such as geodesy, test of fundamental constants, precision navigation, time keeping, and astronomy\cite{ref7,ref8,ref9,ref10,ref11,time}.

Unfortunately, the ultra-precision timescale and frequency signal are affirmatively derived form cumbersome and expensive systems which are normally placed at national metrology institutes or several universities. The status causes a strong motivation to develop effective dissemination systems for distributing these superprecise time and frequency signals.  Among the existing schemes, the optical fiber is regrading as an ideal transmission medium because of its advantages of low power attenuation, high reliability, lower cost, and large communication capacity, which has been fleetly and widely deployed for time and frequency transfer\cite{ref15,ref16,ref17,ref18}. And with it, various transmission techniques for the above-mentioned metrological time and frequency signal are a rapidly developing alternative. For the optical frequency reference, an additional phase noise induced by the variation of fiber physical length and refractive index induced as a result of vibration and temperature fluctuation will deteriorate instability. To cancel the so-called Doppler noise, the active noise cancellation scheme was firstly proposed and implemented by \emph{Ma} \emph{et al.} in 1994, by which the phase error between the local refer signal and the round-trip optical signal is derived and further feedback to the voltage-controlled oscillators\cite{ref12}. For the timescale signal, \emph{Steven R. Jefferts} \emph{et al.} came up with the two-way time transfer technique using optical fibers in 1996, which allowed time transfer over short distances (km) with stabilities less than 10 ps\cite{IEEE97}, then the varies techniques to acquire the delay fluctuation and to compensate for it in the fiber\cite{wangbo,420km,ref19}. Additionally, a dissemination system for the ultra-stable 100 MHz microwave frequency signal along a standard fiber network has been demonstrated by \emph{F. Narbonneau} \emph{et al.} using amplitude modulation of an optical carrier\cite{Narbonneau}.

For some scientific researches, it is a pressing need for the simultaneous dissemination of time and frequency signals, such as VLBI and further redefinition of the second \emph{et al.}\cite{VLBI,redefinition,timescale}. In order to meet this demand, several kinds of schemes are proposed to simultaneously transfer the time and frequency signals over a single long-haul fiber link\cite{same,elstab}. The most possible and convenient solution is to combine the time and frequency signals by using wavelength-division multiplexing (WDM)-based fiber network. In \cite{lopze}, the authors demonstrated an optical frequency reference transfer compatible with data traffic using the internet fiber network with the WDM technique. \emph{Ukasz liwczyski} \emph{et al.} observed very good stability of two delivered signals in a WDM transfer link: Allan deviation approach 4$\times10^{-17}$ (for 10 MHz frequency signal), and time deviation below 1 ps for one pulse per second (1 PPS) time signal both at one day averaging time in 2013\cite{ref21}. Ref\cite{wangbo} also shows an 80 km WDM fiber link for two references (time and 9.1 GHz microwave frequency) synchronization system at the 5$\times10^{-19}$ accuracy level of the frequency, and 50 ps precision level of the time. Besides, the time and frequency of simultaneous transfer research through fiber optics links have demonstrated in recent years, and there are very meaningful results\cite{ref20,ref22,ISI:000361035300116,joint,liuqin,lopez2}. Up to now, the familiar simultaneous transfer of time and frequency references based on the WDM systemfocuses on signals time-microwave or time-optical frequency. Taking into account the multiple needs and the flexible diversity of options for remote users, the WDM-based solution for for offering the common reference signals (time and microwave frequency)  and most precision optical frequency reference at the same time has become important and necessary.

In this work, we present a simultaneous transfer for the optical frequency, 10 GHz microwave frequency, and 1 PPS time signal via a WDM-based fiber link. Thereinto, we use the active noise cancellation method and electronic phase compensation technique to compensate the fiber noise for optical and 10 GHz microwave frequency signals respectively, while the dual-wavelength time synchronization technique is applied for 1 PPS time signal. We observe a relative frequency instability of 3$\times10^{-19}$ for optical frequency and 1.4$\times10^{-17}$ at 10000 s for 10 GHz microwave frequency both at 10000 s integration time and the uncertainty of the 1PPS is 2.08 ps. Note that, the time and frequency signals are all generated from the National Time Service Center (NTSC) lab,  and the three transfer systems are synchronized by using the same standard timebase. This technique provides a flexibly and freely chosen metrology time and frequency signal for remote users. 
\section{The scheme of the WDM-based system}
\label{sec:examples}

The main schematic diagram of the WDM-based system is shown in Fig.1. The bandwidth of every channel in the WDM is about 100 GHz (~0.8 nm) with isolation ratio of 50 dB, and the insert loss is 1 dB. In this setup, optical frequency signal travels in C 34 channel, 10 GHz microwave frequency goes in C 35 channel on the uplink and goes in C 37 channel on the downlink, and 1 PPS time signal uses C 42 and C 43 channels for the remote time synchronization. Otherwise, the other channel can be used as extension path for other communications services. In the local site of this scheme, the five specific wavelengths carring different time or frequency signals are combined into the WDM for traveling in the fiber link. At the remote site, another WDM is used as a de-wave division device to distinguish different signals into their respective channels for each transfer system and to evaluate the measurement results.
All the above-mentioned setups are placed in the NTSC laboratory, and a 50 km spooled fiber link in this experiment is employed. Note that, a dispersion compensation fiber is placed at the output port of the C 33 channel, which is used to compensate mode dispersion of the 10 GHz microwave frequency reference, with an additional insertion loss of 5 dB.
\begin{figure}[t]
	\centering
	{\includegraphics[width=\linewidth]{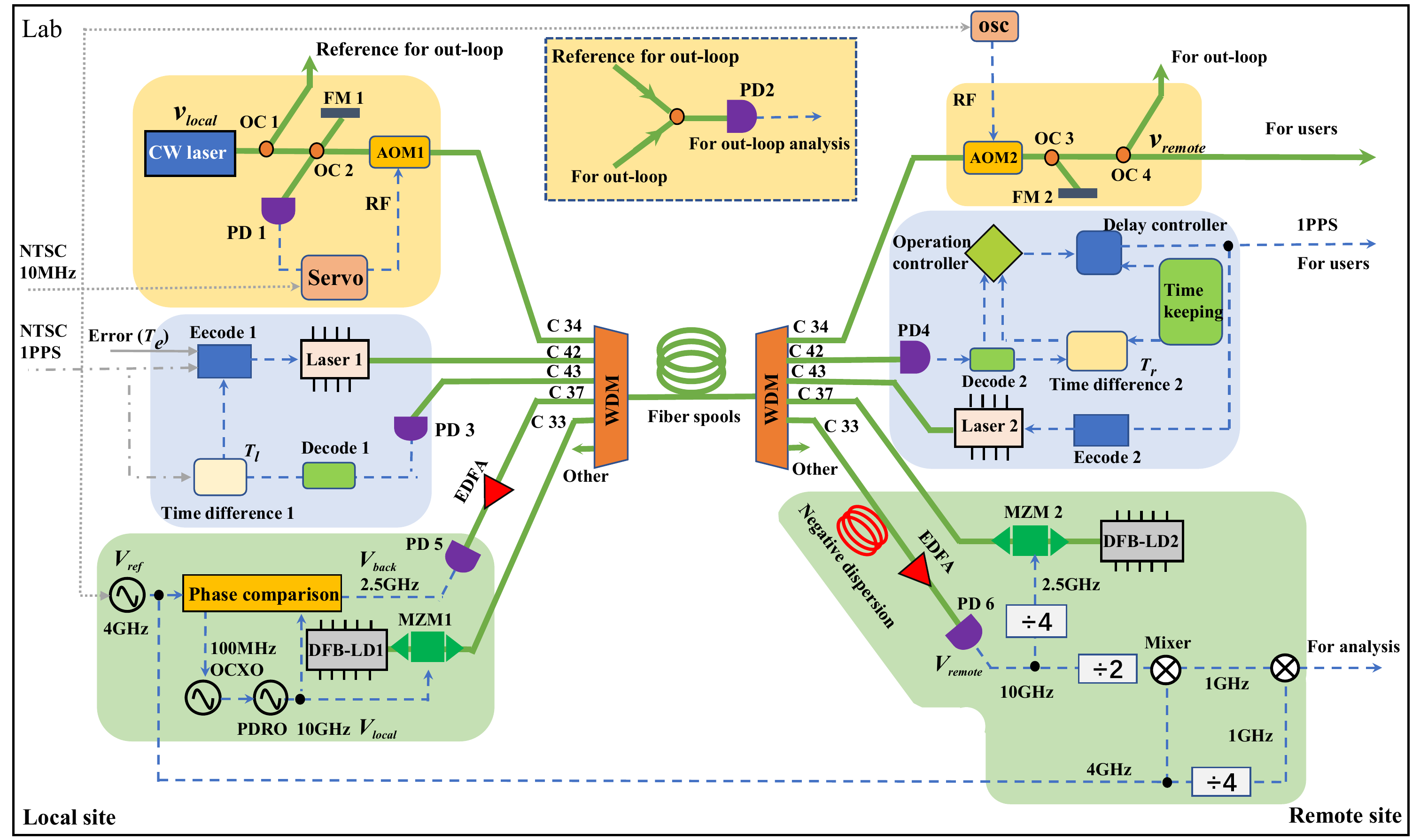}}
	\caption{The principle of optical frequency, 10 GHz microwave frequency, and 1 PPS time transfer in the WDM-based system. WDM:wavelength division multiplexing, AOM: acoustic-optical modulator, OC: optical coupler, PD: photonic detector, FM: faraday mirror, osc: Crystal oscillator, RF: radio frequency, EDFA: Erbium Doped Fiber Amplifier, OCXO: oven-controlled crystal oscillator, PDRO: phase-locked resonant oscillator, MZM: Mach-Zehnder modulator, DFB-LD: distributed feedback laser diode.}
	\label{fig:false-color}
\end{figure}

For coherent optical frequency transfer link (marked on Fig.1 with a yellow background color), the Doppler noise suppression method is used to actively eliminate noise induced by the transmission link to realize the local ultra-stable optical frequency signal copied at the remote site\cite{ref23}. The carrier laser source is an ultra-stable continuous-wave (CW) with a linewidth of 1.9 Hz, which center wavelength is 1550.12 nm\cite{ref24}. The frequency of laser output laser $\upsilon_{local}$  is expressed as
\begin{equation}
	\upsilon_{local}=\frac{d\phi_0(t)}{dt}.
\end{equation}
The laser is split into two parts by a 90/10 optical coupler (OC1), the 90$\%$ part is injected into fiber for transfer, while the 10$\%$ part is used as reference light for an out-loop evaluation of transmission quality. Then the 90$\%$ transferred light is split into two parts by the 50/50 OC2, one part is directly reflected in the photodetector (PD1) by a Faraday mirror 1 (FM1), and the other part passes through acoustic-optical modulator 1 (AOM1). And then the transmitted signal goes into the C 34 channel of WDM for transmission in the fiber. Here the AOM1 shifts the optical frequency signal with 110 MHz and acts as a noise cancellation device. On the remote side, the optical frequency light transfers out from the WDM and passes the AOM2. Then the light is further split into two parts again by OC3, the most part is returned to the PD1 by FM2. While the less part is used for evalution of transfer performance, expressed as
\begin{equation}
	\upsilon_{remote}=\upsilon_{local}+\frac{d[\phi_c(t-\tau)+\phi_{transfer}]}{dt},
\end{equation}
here the $\phi_c(t-\tau)$ is the compensatation phase, $\tau$ is the sigle-trip time, $\phi_{transfer}$ is the transfer phase. Because of the swirling effect of the Faraday mirror, the reference signal and returned signal have the same polarization state to generate a stable beat note signal at PD1. We extract the error signal from the beat note signal and use it to compensate the fiber noise through the AOM1. Then the  $\upsilon_{remote}$ is transformed into
\begin{equation}
	\upsilon_{remote}\propto\frac{d\phi_0(t)}{dt}.
\end{equation}
 The AOM2 helps to discriminate the return light from the stray reflections by splices and connectors, which shift the optical frequency of 50 MHz. PD2 is used to characterize and analyze the performance of out-loop instability.  For reducing the impact on the environment, the fiber length of the interferometer is designed to be as short as possible.

The C 42 channel and C 43 channel are used to transfer the 1 PPS time signal. For 1 PPS time dual-wavelength synchronization transfer, there are also local site and remote site at the system, the schematic diagram as Fig.1 is shown in (marked with a blue background color). The 1 PPS signal from NTSC in the local site is modulated to laser 1 (center wavelength is 1543.7 nm) by the encoder1, then the light travels in C 42 channel to the remote site. The signal is detected by PD4, and decoded for fiber dispersion delay information, then sent to the operation controller. The incoming 1 PPS time signal is compared with the timekeeping 1 PPS signal of remote site module by a time difference measurement module 2, this time difference signal $T_r$ is also sent to the operation controller. The operation controller processes the time difference and error data $ T_e $ caused by fiber dispersion, then uses the delay controller to adjust the 1 PPS signal output by the timekeeping module. The 1 PPS time signal of the remote site is modulated to the laser 2 (center wavelength is 1542.9 nm) by the encoder2 and travels C 43 channel back to the local site. The back signal is detected by PD3 and decoded by decode 1, then the signal is fed to the time difference module 1 for comparison with the 1 PPS time in the local site. According to this scheme, the time delay compensation control amount  $T_d$ of dual-wavelength time synchronization can be expressed as
\begin{equation}
T_d=\frac{1}{2}(T_l-T_r)+T_s+T_e.
\end{equation}
Where the $T_l$ and $T_r$ are measured by the time difference measurement module of the local and remote site. $T_s$ is the system delays and can be calibrated in the equipment. $T_e$ is the time delay caused by the fiber dispersion, and the $T_e$ can be calculated by
\begin{equation}
	T_e=\frac{1}{2}{D}(\lambda_1-\lambda_2)L,
\end{equation}
here the $D$ is dispersion coefficient, the $\lambda$ is laser wavelength, and the $L$ is fiber length.

In the 10 GHz microwave frequency transfer link, there are also local and remote ends located in our laboratory (marked on Fig.1 with a green background color). The reference frequency $V_{ref}$ is 4 GHz generated by an RF source (Keysight E8257D) which is synchronized with the NTSC 10 MHz timebase, expresses as
\begin{equation}
	V_{ref}\propto\sin(\omega_rt+\phi_r).
\end{equation}
 At the local site, the transferred 10 GHz microwave frequency $V_{local}$ is generated from a phase-locked resonant oscillator (PDRO), which is locked to a low noise 100 MHz oven-controlled crystal oscillator (OCXO). At the remote site, the transferred 10 GHz microwave frequency $V_{remote}$ is detected by PD6 and then is divided to 2.5 GHz by a frequency divider. This 2.5 GHz signal is used as the backward signal. The intensity of the distributed feedback laser diode (DFB-LD) at 1547.7 nm for the forward link and 1550.9 nm for the backward link are modulate by forward and backward microwave frequency signals with Mach-Zehnder modulator (MZM), respectively. Hence, the forward signal travels in C 37 channel and the backward signal travels in C 33 channel. The 2.5 GHz backward signal is detected by PD5 at the local site and the signal $V_{back}$ carries the roundtrip noise over the fiber link. Based on $V_{remote}$, $V_{ref}$, and $V_{back}$, the phase comparison system gives the error control signal which is fed to 100 MHz OCXO to cancel the fiber link noise. Then the $V_{local}$  is transformed and expressed as
\begin{equation}
	V_{local}\propto\sin(\frac{5}{2}\omega_rt+\frac{5}{2}\phi_r+\omega_0\tau),
\end{equation}
where $\tau$ is the one-way trip propagation delay of fiber. And the  $V_{remote}$ is
\begin{equation}
	V_{remote}\propto\sin\frac{5}{2}(\omega_rt+\phi_r).
\end{equation}
 The effect of the chromatic dispersion in the fiber will deteriorate microwave frequency transmission performance over the link. To reduce the effect of the chromatic dispersion (~17 ps/nm·km) of fiber link on microwave frequency transfer, a large negative dispersion fiber (~-950 ps/nm) is set in the microwave frequency transfer system.

\section{The experimental results and discussion}

To evaluate the performance of this WDM-based time and frequency transfer system. The results are performed in three parts, optical frequency transfer, 1 PPS time signal transmission, and 10 GHz microwave frequency transfer.

\subsection{Optical frequency transfer}
\begin{figure}[h]
	\centering
	{\includegraphics[width=\linewidth]{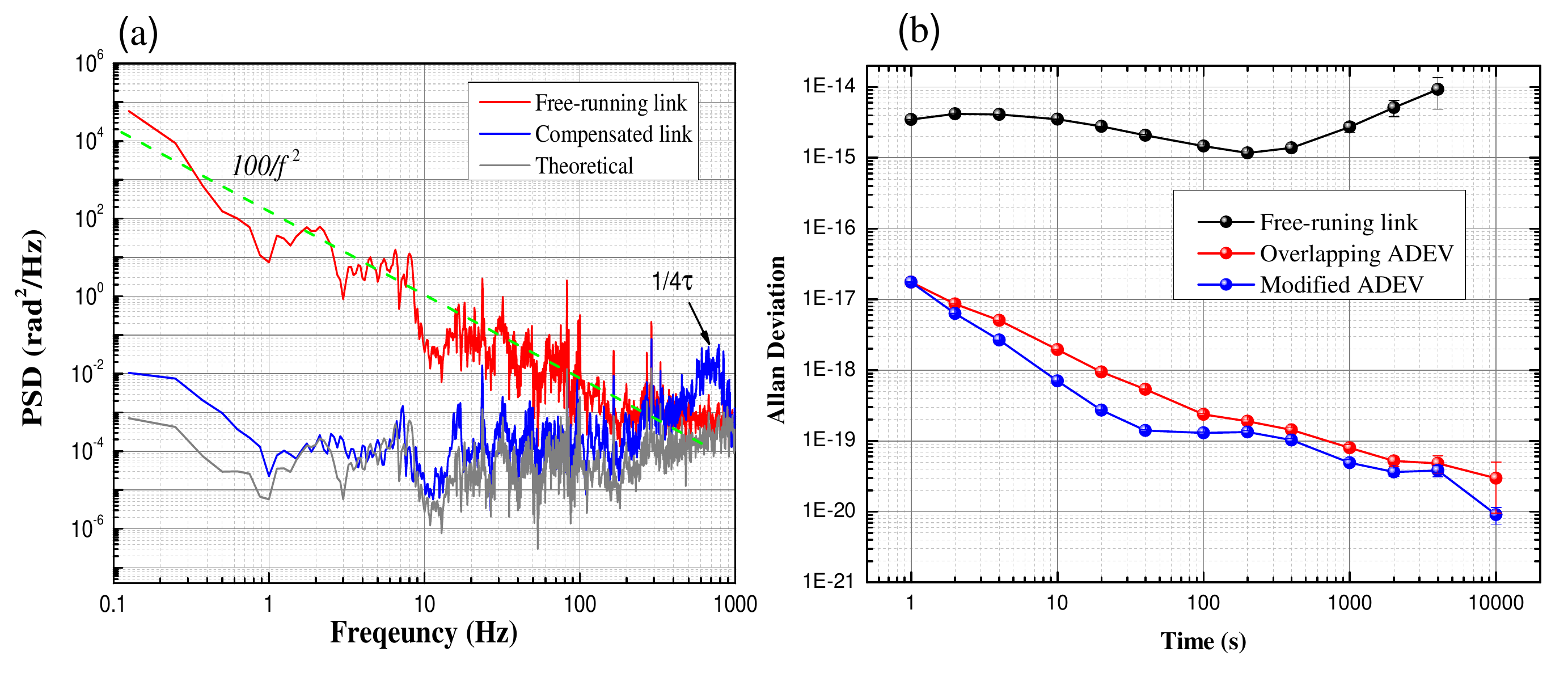}}
	\caption{ (a) the phase noise of the optical frequency transfer, the red curve is the phase noise of the free-running link, the blue curve is the phase noise of the locked and compensated link, and the gray curve is the phase noise of the theoretical value; (b) the Allan deviation of the optical frequency transfer, the black dot curve is the free-running link, the red dot curve is the overlapping ADEV of the compensated link, and the blue dot curve is the modified ADEV of the compensated link.}
	\label{fig:false-color}
\end{figure}
The phase noise PSD of the optical frequency transfer link is measured with a fast Fourier transform spectrum analyzer (Stanford Research System, SR785) as shown in Fig.2(a). The red line is the free-running link and it exhibits a relationship with frequency as $100/$\emph{f}$^2$ in the 0.1-1000 Hz\cite{ref13}, which indicates a white phase noise disturbance along the fiber link. The blue line is the phase noise of the compensated link, it has more than 5 orders below than free-running link at 1 Hz. The higher peak in 600 Hz is caused by the servo control bandwidth limited of 1/4$\tau$\cite{ref13}. According to the theoretical noise reduction ratio $1/3(2$$\pi$$\tau$\emph{f}$ )^2$, with $\emph{f}$ being the Fourier frequency, and $\tau$ being the single-trip delay time\cite{ref13,ref25}, the corresponding residual phase noise is calculated as shown by the grayline. The phase noise of stabilized link is coincident with the gray line which illustrates that the loop control bandwidth is the main reason limiting the system performance.

For the coherence optical frequency transfer system, a no-dead-time frequency counter (FXE K+K) is employed to record the out-loop beat-notes detected by PD2 (see Fig.1 dash line little window), with a gate time of 1 s operating in $\Lambda$-type mode. The fractional instabilities of the optical frequency with and without compensation are shown in Fig.2(b). For the free-running link, the overlapping Allan deviation is 3.5$\times10^{-15}$ at 1s averaging time and and decrease to 9.2$\times10^{-15}$ at 4000 s averaging time. When the phase noise is canceled, the overlapping Allan deviation with the black square line is 1.8$\times10^{-17}$ at 1 s averaging time and scales down as 1/$\tau$ slop to 3$\times10^{-19}$ at 10000 s averaging time. The modified Allan deviation (the red dot line) is also calculated, the 1 s instability is the same as the overlapping mode, while it reach 9$\times10^{-20}$ at the 10000 s averaging time.

\subsection{1 PPS time signal transfer}
\begin{figure}[h]
	\centering
	{\includegraphics[width=9cm,height=12cm]{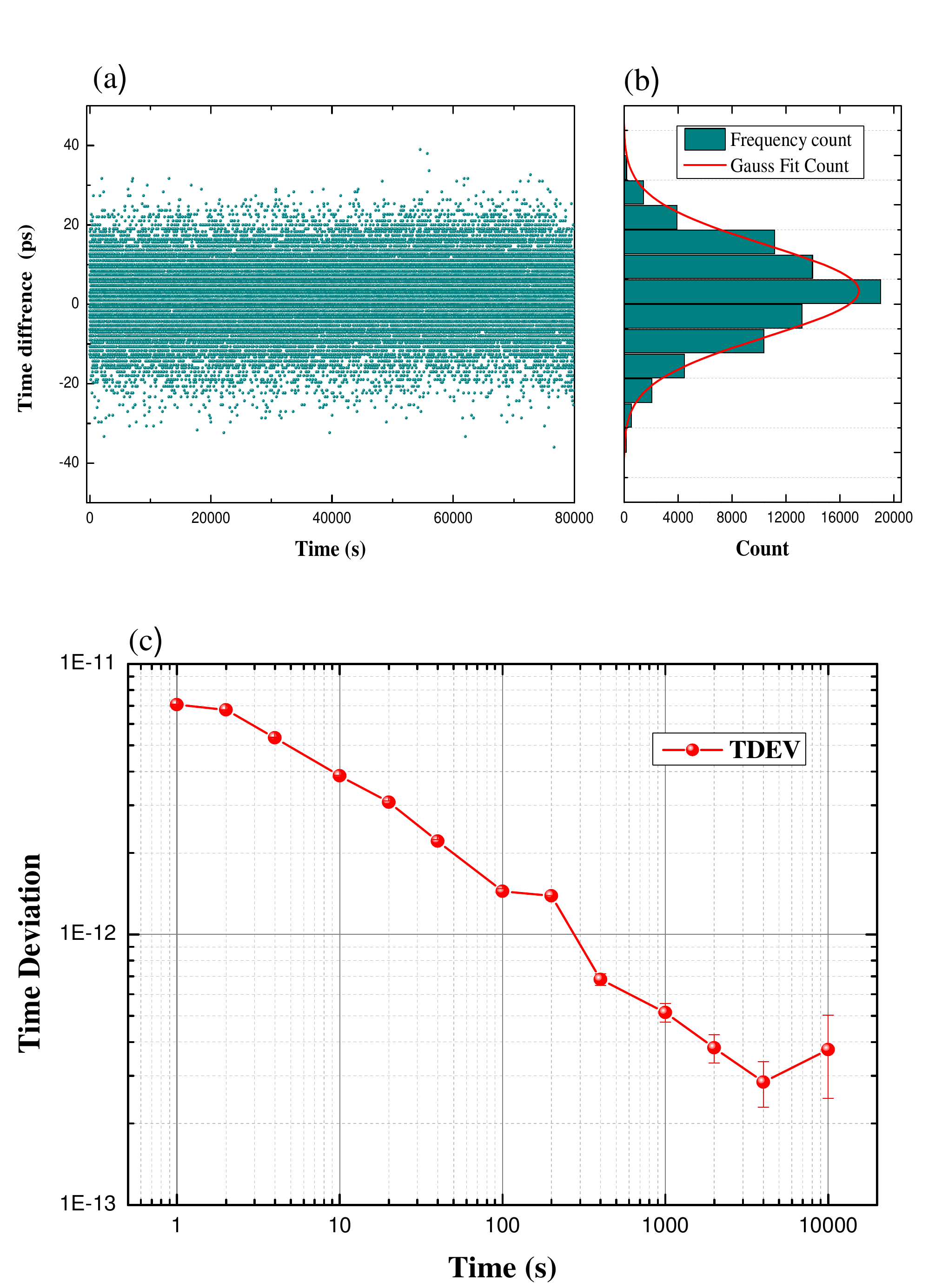}}
	\caption{(a) the time interval varition of the 1 PPS time signal;
		(b) the statistics results of the I PPS time data; (c) the TDEV of the 1 PPS time transfer.}
	\label{fig:false-color}
\end{figure}
In order to test the performance of the 1 PPS time signal transfer through C 42 and C 43 channels in the WDM-base system, the transmitted 1 PPS time signal and the local reference 1 PPS are compared.

When the local site and remote site are working properly, the 1 PPS signal from remote site output and the reference 1 PPS are compared by a time difference measure device (SR620) with a gate time of 1 s. During the test, the errors and time delays in the system such as dispersion error and time delay of the difference in cable length have been calibrated by the time transfer equipment. Fig.3(a) shows the time difference data in one day (the green dots), in which the peak-to-peak is about 70 ps, the time uncertainty is calculated to be 2.08 ps, and the time data follows the gauss fit (the red line in Fig.3(b)).

At the same time, the time synchronization results of the 1 PPS time signal are shown as time deviation (TDEV) in Fig.3(c). It can be found that TDEV is 7.06$\times10^{-12}$ at 1 s integration time and reaching 3.75$\times10^{-13}$ at 10000 s integration time (the red dot line). Because of the temperature fluctuation by the air conditioner in the lab (approximately 16 minutes as a cycle) and the wavelength shift of the lasers, the long-term time instability at 200 s deteriorate in the curve.

\subsection{10 GHz microwave frequency transfer}
\begin{figure}[h]
	\centering
	{\includegraphics[width=\linewidth]{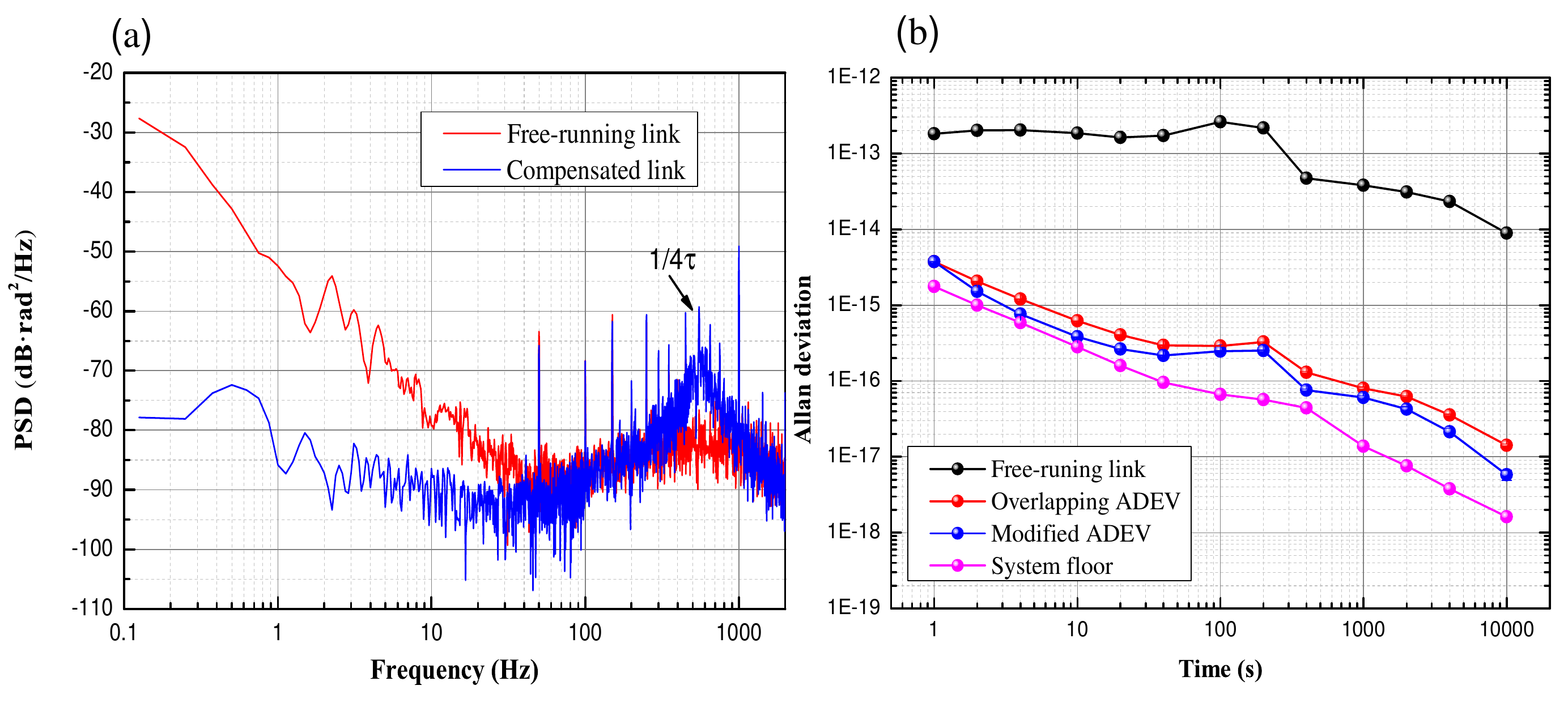}}
	\caption{(a) the phase noise of the 10 GHz microwave frequency transfer, the red curve is the phase noise of the free-running link, the blue curve is the phase noise of the compensated link; (b) the Allan deviation of the 10 GHz microwave frequency transfer, the black dot curve is the free-running link, the red dot curve is the compensated link overlapping ADEV, while the blue dot curve is the modified ADEV, the pink dot curve is the 1-m tranfer link system floor}
	\label{fig:false-color}
\end{figure}

For evaluating the performance of 10 GHz microwave frequency transfer in the WDM-based system, we measure the phase difference between the local reference signal $V_{ref}$ and the received signal $V_{remote}$ in the locked loop (see Fig.1).
The phase difference of $V_{ref}$ and $V_{remote}$ is down-converted to direct current (DC) voltage $V_{transfer}$ and measured by a multimeter (Keysight 3458A). Fig.4(a) shows the phase noise of the microwave frequency transfer link at 10 GHz. The red curve is the free-running link measured by mixing $V_{local}$ and $V_{remote}$. The blue curve represents the compensated link phase noise obtained by measuring the $V_{transfer}$. The peak of the compensated link at about 600 Hz is the servo bandwidth of the 10 GHz microwave frequency transfer which agrees with Ref.\cite{ref13}. It can be analyzed from the phase noise curves that the noise has been canceled 30 dB in 1 Hz by locking the transfer link.

Fig.4(b) shows the instability of the 10 GHz microwave frequency calculated from the link delay of the free-running and compensated link. The instabilities of overlapping Allan deviation as red dot curve is 4.0$\times10^{-15}$ at 1 s integration time and reaching 1.4$\times10^{-17}$ at 10000 s integration time. A blue dot curve also shows the modified Allan deviation of the stablized link which achieves  4.0$\times10^{-15}$ at 1 s integration time and decreases to 5.5$\times10^{-18}$ at 10000 s integration time. The black dot line is plotted to the uncompensated link as the comparison with the compensated link. At the same time, the system floor of the 1 m compensated link is shown in the figure as a comparison with the pink dot curve. It can be found that there are bumps in the curves, which is also because the temperature disturbances in the laboratory by the air condition.

\subsection{Simultaneous and independent transfer comparison}
For comparing the performance of the WDM-based simultaneous transfer system with independent transfer via the link, an independent experiment of these time and frequency signals transfer is also demonstrated in this link. In the experiment, the individual measurement of a specific system is evaluated by blocked the signal of other channels to ensure that there is only one signal in the transfer link. 

The comparison results are shown in table 1 (ST refers to simultaneous transfer, IT refers to independent transfer). Here, the overlapping ADEV is used to characterize the transfer performance of the optical frequency,and 10 GHz microwave frequency, while the 1 PPS time signal is expressed by uncertainty. For optical frequency signal, the short-term (1 s) simultaneous and independent transfer are 2.12$\times10^{-17}$ and 2.15$\times10^{-17}$, while the long-term (4000 s) are 2.62$\times10^{-20}$ and 2.18$\times10^{-20}$. The transfer results in the two cases are almost equal. The same situation also occurs in the 10 GHz microwave frequency and the 1 PPS time signal. The WDM-based simultaneous transfer system has shown a similar performance compares to the independent transmission system, which proves its usefulness and capability in time and frequency related applications. However, when the optical amplifiers are added to the link in order to amplify the signals, there will be new problems, such as the distribution of signal gain and influence between channels, etc, which will be our further investigation.
\begin{table}[h]
	\centering
	\caption{\bf comparison of the simultaneous transfer and independent transfer}
	\begin{tabular}{ccccc}
		\hline
		    $ signal\ type $      &     $\  ST\ (1s)$      &      $\ IT\ (1s)$      &    $\  ST\ (4000s)$    &    $\ IT\ (4000s)$     \\ \hline
		                     &  \\
		 $Optical\ Frequency$  & $2.12\times10^{-17}$ & $2.15\times10^{-17}$ & $2.62\times10^{-20}$ & $2.18\times10^{-20}$ \\
		                     &  \\
		$10\ GHz\ microwave\ Frequency$ & $5.66\times10^{-15}$ & $5.67\times10^{-15}$ & $3.26\times10^{-17}$ & $3.25\times10^{-17}$ \\ \hline
		                     &  \\
		$1\ PPS\ (uncertainty)$ & $ST=2.08\ ps$     &          $$          & $IT=2.09 \ ps$     &  \\ \hline
	\end{tabular}
	\label{tab:shape-functions}
\end{table}

\section{Conclusion}

In summary, we demostrate a WDM-based system for simultaneous transfer of the optical frequency, 10 GHz microwave frequency, and 1 PPS time signal over a 50km
spooled fiber link, which are synchronized with the national standard time or frequency from NTSC. The solution, combining the three systems in one fiber, allows the users choose any reference signals according to their application scenarios. Additionally, the WDM-based system can be flexibly integrated with the existing communication network without the need of a dedicated optical fiber. After being compensated, the fractional frequency instability of optical frequency is 1.8$\times10^{-17}$ at 1 s and 3$\times10^{-19}$ at 10000 s integration time, and 4$\times10^{-15}$ at 1 s integration time and 1.4$\times10^{-17}$ at 10000 s for 10 GHz microwave frequency transfer. For 1 PPS time signal transfer, the uncertainty is calculated to be 2.08 ps and the instability is 7.06 ps at 1 s averaging time.
These results show the WDM-based system has a great potential for the time and frequency transfer with commercial fiber link, so as to save costs and realize nationwide time and frequency transmission in the future.

In the next work, we are planning to conduct longer-distance transmission with bidirectional Erbium-doped optical amplifiers in communication fibers. For testing and improving the system reliability, it is also necessary to study interference between channels and the gain distribution of optical amplifiers.

\begin{flushleft}
	\textbf{Acknowledgment:} The authors would like to thank W. Zhao  for technical assistance. The authors acknowledge the Strategic Priority Research Program of the Chinese Academy of Sciences (Grant No. XDB21000000) and the Open Project Fund of State Key Laboratory of Transient Optics and Photonics, Chinese Academy of Sciences(Gtant No. SKLST202011).	
\end{flushleft}

\begin{flushleft}
\textbf{Conflicts of Interest:} The authors declare no conflict of interest.	
\end{flushleft}

%%%%%%%%%%%%%%%%%%%%%%%%%%%%%%%%%%%%%%%%%%
\end{paracol}
\reftitle{References}


\externalbibliography{yes}
\bibliography{template}

%%%%%%%%%%%%%%%%%%%%%%%%%%%%%%%%%%%%%%%%%%
%% for journal Sci
%\reviewreports{\\
%Reviewer 1 comments and authors’ response\\
%Reviewer 2 comments and authors’ response\\
%Reviewer 3 comments and authors’ response
%}
%%%%%%%%%%%%%%%%%%%%%%%%%%%%%%%%%%%%%%%%%%
\end{document}